\pdfoutput=1 
\documentclass[12pt,preprint]{aastex}
\begin{document}
\normalsize

\shorttitle{High-Velocity Cloud A0}
\shortauthors{Verschuur}

\title{High-Resolution Observations and the Physics of High-Velocity Cloud A0}

\author{Gerrit L. Verschuur}

\affil{Physics Department, University of Memphis, Memphis, TN 38152}

\email{verschuur@aol.com}

\begin{abstract}
The neutral hydrogen structure of high-velocity cloud A0 (at about $-$180 km s$^{-1}$) has been mapped with a 9.\arcmin1 resolution.  Gaussian decomposition of the profiles is used to separately map families of components defined by similarities in center velocities and line widths.  About 70\% of the HI gas is in the form of a narrow, twisted filament whose typical line widths are of order 24 km s$^{-1}$.  Many Òbright features with narrow line widths of order 6 km s$^{-1}$, clouds, are located in and near the filament.  A third category with properties between those of the filament and clouds appears in the data.  The clouds are not always co-located with the broader line width filament emission as seen projected on the sky.  Under the assumption that magnetic fields underlie the presence of the filament, a theorem is developed for its stability in terms of a toroidal magnetic field generated by the flow of gas along field lines.  It is suggested that the axial magnetic field strength may be derived from the excess line width of the HI emission over and above that due to kinetic temperature by invoking the role of Alfv\'en waves that create what is in essence a form of magnetic turbulence.  At a distance of 200 pc the axial and the derived toroidal magnetic field strengths in the filament are then about 6 $\mu$G while for the clouds they are about 4 $\mu$G. The dependence of the derived field strength on distance is discussed.  
\end{abstract}


\section{Introduction}
The goal of this work is to report on the results of the high-resolution (9.\arcmin1) HI mapping of high-velocity feature A0 (also known as HVC 132+23-211) and a small segment of AI, and to offer a theory to account for the structure found in the study. The data were obtained using the Robert Byrd Green Bank 100-m telescope (GBT) of the National Radio Astronomy Observatory (NRAO).\footnote{The National Radio Astronomy Observatory is a facility of the National Science Foundation operated under cooperative agreement by Associated Universities, Inc.}

Observations of high-velocity cloud Complex A and its individual concentrations goes back at least 40 years.  The first relatively high-resolution observations of the Complex A were made by Giovanelli, Verschuur \& Cram (1973) using the late 91-m radio telescope of the NRAO with a beam width of 12\arcmin\ but in retrospect the data were not very sensitive and were subject to sidelobe contamination. They mapped the main concentrations and labeled them AI through AV. Subsequently Verschuur, Cram \& Giovanelli  (1972), again using the 91-m telescope, obtained limited data for A0. The bulk of the HI emission from HVCs was mapped by Hulsbosch (1975).  His survey included Complex A and noted the presence of the feature A0. 

Davies et al. (1976) used the 91-m and 42-m NRAO telescopes to map AIV and several other anomalous velocity HI features and found angular structure on beam-width scales and emission profiles with narrow-velocity widths comparable to those noted by Verschuur, Cram \& Giovanelli  (1972), of order 5 $-$ 7 km s$^{-1}$ (full-width at half maximum).  Schwarz, Sullivan \& Hulsbosch (1976) published aperture synthesis observations of A0 using 2\arcmin\ \& 4\arcmin\ angular resolutions but with a very large (27 km s$^{-1}$) bandwidth.  They were able to identify concentrations with angular extents of order 5\arcmin\ and found evidence for filamentary structure as well.  The detailed structure of A0 (also known as HVC 132$+$23$-$211) was mapped with an angular resolution of 9\arcmin\ by Hulsbosch (1978) who described the main feature as a "small isolated cloud"  and suggested that the apparent co-existence of both narrow (5 km s$^{-1}$ wide) and broad components (from 10 to 15 km s$^{-1}$ wide) indicated that the HI had a two-component structure.  Wakker \& Schwarz (1991) argued for a similar core-halo model and many others have embraced the idea.  Wolfire et al. (1995) showed that the lower galactic halo pressure of the order of 4,000 cm$^{-3}$ K (log(nT) $=$ 3.6) provides the external pressure to confine the HI in the core-halo model.  Schwarz \& Wakker (2004) offered a similar argument.  (Note that the area covered by both Verschuur, Cram \& Giovanelli  (1972) and Hulsbosch (1978) included only the southern-most features in A0 as compared to the present study.)

High angular resolution (1\arcmin) aperture synthesis observations of several of the features in Complex A were published by Schwarz \& Oort (1981) for AI and by Wakker, Vijfschaft \& Schwarz (1991) for AIV.  Again, angular structure of several arcminutes in extent was found and velocity structure on a scale of a few km s$^{-1}$ was recognized in the emission profiles. 

While the feature A0 has usually been classed as an example of a high-velocity cloud its true morphology was not really known. Here it is found to be composed of a dominant filament and several dozen small angular scale clouds.  The filamentary nature of diffuse interstellar HI has been widely recognized for years; see, for example, Colomb, P\"oppel, \& Heiles (1980), Verschuur (1991a, 1991b), Verschuur et al. (1992), and Burton, Hartmann \& West (1996) and the all-sky data of Hartmann \& Burton (1997). More recently, in a catalog of southern hemisphere, anomalous-velocity HI based on the HI Parkes All-Sky Survey (HIPASS) data, Putman et al. (2002) report that most high-velocity emission features have a filamentary morphology and are loosely organized into large complexes extending over tens of degrees. 

In the first several sections of this paper the data are prepared so that a theory for filament stability can be developed and applied.  The data are discussed in \S2 and the results of a Gaussian analysis of the observed profiles are presented in \S3.  In \S4 the values of the crucial parameters, distance and kinetic temperature, required to convert observable quantities into those required for a quantitative analysis of the data are discussed.  In \S5 a comprehensive theory invoking the role of magnetic fields to account for the filament and the clouds is presented.  The theory requires knowledge of the axial magnetic field strength and it is hypothesized in \S6 that this may be obtained from the broadening of the observed HI profiles by Alfv\`en waves.  The theory is applied to the data in \S7 and conclusions are summarized in \S8.

\section{The data}
A complete mapping of A0 between longitude limits 129.\arcdeg7 and 138.\arcdeg5 and latitudes 22.\arcdeg8 and 28.\arcdeg5 was carried out using the GBT.  A channel bandwidth of 0.0583 km s$^{-1}$ was used and in this report HI emission between velocity limits of $-$250 and $-$50 km s$^{-1}$ was examined. The r.m.s. noise level on the data is of order 0.1 K in one channel. Emission profiles were obtained every 3.\arcmin5 which means that the data were fully sampled given that the beam width is 9.\arcmin1 in the L-band.  

To make maps of the morphology of the HI structures raw antenna data were used and re-sampled four times and then the contours were drawn.  This allows sharp (pixel-like) edges to be smoothed out in the process.  When required to calculate numerical values of peak column density the observed antenna temperatures were converted to brightness temperatures using a small correction (0.94) for beam efficiency of the GBT. In what follows the results of the mapping of A0 will be shown in several different coordinate systems to illustrate how each highlights a different aspect of the data.

The morphology of the A0 area as observed with the GBT is shown in Figure 1.  The first three plots show position-position contour maps of integrated HI brightness in the velocity intervals noted in the caption.  This type of display leads to the concept of cloudiness in the medium.  Fig. 1d is an inverted gray scale image of the HI emission integrated between $-$220 to $-$120 km s$^{-1}$, which encompasses most of the velocity range of the associated HI.  It clearly reveals that a twisted filament links several areas of enhanced emission where the so-called cloudiness is present.  

Figure 2 shows examples of HI brightness contour maps as a function of position and velocity at three closely spaced latitudes.  The presence of a wave-like or sinusoidal-like velocity pattern can be recognized at high negative velocities.  As this feature is followed through {\it l,b,v} phase space its course can be followed from map to map.  Similar {\it l,v} maps were produced at every latitude at which data were obtained.  If the velocity were somehow related to distance then this feature would have a spatial helical component, but that assumption cannot be readily justified.  There is a strong hint in the data that the presence of the anomalous velocity gas at velocities greater that $-$120 km s$^{-1}$ in Figs. 2a \& b shows an associated disturbance around -85 km s$^{-1}$.  

Figure 3 shows the {\it b,v} plot of the averaged HI brightness over the longitude range {\it l} $=$136.\arcdeg3 to 137.\arcdeg3 and it illustrates a few key points that emerge from even this cursory glance at the data.  The first is that while area maps such as Fig. 1 conjure up visions of cloudiness, Figs 2 \& 3 suggest greater complexity.  In the upper part of Fig. 3a a tongue of HI emission associated with the filament sweeps across the map and at its high negative velocity tip a distinct cloud is found as well as three others that appear separated in velocity and position.  It will be shown below that the line widths typically associated with the filamentary gas are large, of order 24 km s$^{-1}$, whereas for the clouds they are of order 6 km s$^{-1}$. This is the first indication that a model of high-velocity gas consisting of a two-component core-halo structure fails.  Here three of the cold cores, the clouds, are not located in the same direction as the filament (broad line width) emission.  

If the study of this tongue of emission and its immediate vicinity had been limited to this area of sky it would be easy to conclude that the phenomenon seen in Fig. 3 is a manifestation of a so-called velocity bridge or a head-tail structure in high-velocity clouds.  Pietz et al. (1996) found multiple examples of such high-velocity bridges, and Br\"uns et al. (2000) described observations of what they referred to as head-tail high-velocity clouds that have a similar appearance.  Note that the head of the tongue of emission in Fig. 3a at {\it b,v} $=$27.\arcdeg5, $-$190 km s$^{-1}$ manifests a bright small angular diameter peak whose presence fits the head-tail description for this tongue of emission very well.  

In the lower part of Fig. 3a a tongue of emission emerges from the HI gas at velocities around $-$100 km s$^{-1}$.  At first glance this would appear to be another example of the velocity gradient in the filament but it is in fact an indication of rotation about the filament axis.  This is illustrated in the inset in Fig 3a that shows a segment of filament taken from Fig. 1c and the feature is given the name ARC.  The {\it b,v} plot covers the longitude range shown in the inset by the two vertical lines. Thus the tongue of emission in the lower part of the main diagram shows that ARC manifests a very steep velocity gradient across its width of 109 km s$^{-1}$ in one degree of latitude.  This appears to indicate rotation about the axis of this segment of the filament.  Examination of Fig. 3a suggests that ARC connects to the emission from so-called intermediate-velocity HI at {\it v} $>-$100 km s$^{-1}$ as can be inferred by the behavior of the low level contours at $-$80 km s$^{-1}$ at {\it b}$=$25.\arcdeg7.  This may of course indicate no more or less than that the velocity extent of ARC just happens to overlap with that from intermediate velocity HI emission in this part of the sky.  

Fig. 3b \& c show {\it l,v} plots of two clouds, to be discussed below, in order to illustrate the extent to which they either are or are not attached to broad emission associated with the filament.  In Fig. 3b very weak emission links the cloud to the broad emission defining the filament but in Fig. 3c no such connection is found.  Fig. 3d is a {\it b,v} plot at a single longitude of the tip of the tongue of emission at the top of Fig. 3a.  Very weak emission does appear to link the cloud to the filament gas.   

\section{Gaussian analysis of the HI profiles}
In order to begin to understand the physics underlying the HI emission structure of A0 a Gaussian decomposition of the profiles into their components was performed using the algorithm described by Verschuur (2004).  Emission profiles were processed on a case-by-case basis to make sure that the algorithm did not settle on solutions that were seriously different from adjacent profiles separated by only 3.\arcmin5.  By examining the results in the context of the surrounding profiles the small-scale structure in the Gaussian components could be accurately mapped.  As a result, some 5724 Gaussian components were gathered into distinct families defined by similarity in line width and center velocity. This list included only those Gaussians with column densities greater than 1.0 x 10$^{18}$ cm$^{-2}$ to minimize the effect of Gauss fitting noise.  

Figure 4 offer six examples of Gauss fits in order to illustrate several key points.  In each frame the coordinates of the profile position are given and these may be located in Fig. 1.  The properties of the Gaussian(s) fit are indicated under peak temperature, T$_{max}$ in K, center velocity V$_{c}$ and line width W, the full width at half maximum, in km s$^{-1}$ and total column density N$_{H}$ in units of 10$^{18}$ cm$^{-2}$.

Figs. 4a \& b are two directions toward the filament where only the broad component is present.  The increase in the emission at velocities great than $-$120 km s$^{-1}$ in (a) extends into the intermediate velocity regime to blend with HI at $-$80 km s$^{-1}$ and may not be part of A0.  This feature can also be recognized in Fig. 2b between {\it b} $=$23.\arcdeg5 \& 24.\arcdeg0.  In Fig. 4b it appears that a weak component might be present at about $-$165 km s$^{-1}$ with an amplitude of 0.15 K barely great than the noise level in the data but adjacent profiles 3.\arcmin5 away do not show this putative signal.  It is probably indicative of the very low level (non-noise) features found in most of the profiles as can be seen in the other examples in Fig. 4.  Fig. 4c is a profile toward Cloud 2 (to be discussed below) and it shows no obvious broad component of order 24 km s$^{-1}$ wide that defines the filament signatures seen in Figs. 4a \& b.   A profile toward Cloud 3 is shown in Fig. 4d and here two narrow components define the emission line with no evidence of a broad underlying component.  In contrast, Fig. 4e is a profile toward Cloud 5 that shows the filament signature, here 19.8 km s$^{-1}$ wide, and emission from a cold cloud. Fig 4f is a profile toward a feature called PairN(orth),which shows emission from several cold clouds as well as a weak broad component whose line width of 16 km s$^{-1}$ is typical of this area in Fig. 1.  This is consistent with the result of Hulsbosch (1975) who mapped only this area of A0.  Note that in the subsequent grouping of Gaussian components into families, those with a peak temperature less than 0.10 K were rejected so that this feature barely meets the criterion for inclusion.  A possible weak interference signal is the likely cause of the odd structure at $-$226 km s$^{-1}$.  A low-level interference spikes appears present in Fig. 4b at $-$113 \& $-$180 km s$^{-1}$.

A histogram of all the Gaussian line widths ${\bf W}_{obs}$ found in the Gauss fitting is shown in Fig. 5.  Two distinct peaks are associated with clearly different morphological structures.  The broad line width features about 18 $-$ 35 km s$^{-1}$ wide are found to be associated with the pervasive filament while the narrow line width components from 4 to 7 km s$^{-1}$ wide are the hallmark of the compact clouds.  This is in close agreement with the results of Cram \& Giovanelli (1976) whose study of several high-velocity features revealed two prominent line width regimes around 7 \& 23 km s$^{-1}$ wide.  The data for A0 also reveal the existence of several distinct features with line widths between these two prominent values of order 12 km s$^{-1}$.  Toward the central feature in Fig. 1d the profiles are so complex that the Gaussian decompositions were ambiguous from point to point and it was impossible to obtain a reliable sorting into families.  Thus the data for that area are not included in the final analysis.

In an extensive aperture synthesis study of high-velocity clouds, Wakker \& Schwarz (1991) noted that only about 30\% of the total single-dish flux is recovered. This implied that in addition to the bright, narrow features that were prime candidates for aperture synthesis mapping, an extended, warm background must also be present.  The new A0 data confirm this division.  Fig. 5 shows a minimum in line widths at 14.5 km s$^{-1}$ and the sum of all the observed column densities in the narrower line width features (clouds) is 30.8\% of the total column density of all the components, while 69.2\% resides in the broad components (the filament) which is essentially identical to the division reported by Wakker \& Schwarz (1991).  

The clearly identified families of Gaussian components found in the analysis were used to make maps of column density, center velocity and line widths of each family.  In some cases, for example where two narrow line width features were closely spaced on the sky, this step of making maps revealed where the results for a given direction had been assigned to the wrong family and such cases were easily resolved.  Gaussian decompositions were not repeated in order to make certain maps ÒlookÓ better because such a step would introduce subjective decisions.  For the purposes of this report it is the column density maps of the various component families that are discussed although maps of center velocities and line widths were also produced.  

\subsection{Observed properties of the broad line width filament segments}
Several filament segments were fully mapped and their observed properties are summarized in Table 1. Col. 1 gives the working name for the segment.  The number in parentheses indicates segments where two closely spaced cuts were used to estimate the width, which was determined using the 10\% peak column density contour on either side of the feature.  The coordinates of the center of the cuts are given in Cols. 2 \& 3.  Col. 4 is the average center velocity for the feature (one sigma errors) in the mapped area and Col. 5 gives the average line width, ${\bf W}_{obs}$.  Col. 6 gives the apparent kinetic temperature implied by the line width.  The peak column density observed within a given family is given in Col. 7.  Col. 8 gives the angular width of the filament cross section.  The profiles of the integrated segment of ARC shown in Fig. 1c were also decomposed into Gaussians and the average line width is 30.5 $\pm$3.5 km s$^{-1}$.  Given that the velocity gradient in the feature would introduce 16 km s$^{-1}$ of line broadening in the observing beam width, the implies an intrinsic line width of the HI emission of 25.0 km s$^{-1}$, corresponding to the line width regime for other segments of the filament summarized in Fig. 4.  Other mapped filament segments do not show obvious velocity gradients across their width that could be interpreted as evidence for rotation about an axis.  The last line in Table 1 list the average properties of the key parameters.

Figure 6 presents a compilation of area maps for segments of the main filament with the average center velocities and line widths of segments of the filament indicated.  The contours of column density for the broad line width components are overlain on an inverted gray scale map of narrow line width components in the central area where the two overlap. The inset shows a segment of filament that emerges from the bright central feature in the area map in Fig. 1d and its properties are also summarized in Table 1.  The cuts used to determine the filament width are shown in the inset.      
    
Another aspect of the outline of the filament in Fig. 6 that merits comment is that in one area the average line width is larger (29.8 km s$^{-1}$) than any of the others.  The morphology of that area shows an elongated peak along the axis of the filament.  Upon a re-examination of the profiles and the Gaussian components in this direction, it appears that the profiles could equally well have been fit with two slightly narrower, yet still broad, components, but to do so in retrospect would not be consistent with the protocol discussed above.  

Finally, as regards the combination of structures shown in Fig. 6, the filament appears to be very narrow in the area where the arrow shown has been labeled "overlap."  There two broad line width features at distinctly different velocities, labeled Cswy (for Causeway) in Table 1, just overlap.  The emission profiles in that area allow for a clear separation into two broad line width components, Cswy B \& C in Table 1, at velocities of  $-$171 \& $-$149 km s$^{-1}$ while the southern one is seen as a continuation of another broad line width component at $-$135 km s$^{-1}$, which forms an extension of the Bar 2 area ($-$140 km s$^{-1}$) listed in Table 1.  These small segments then join smoothly to the structure at the top-right of Fig. 6, which shows the morphology of the feature called Outlier. A cloud not shown here overlaps this filament component.  These data reveal that while the major filament segments appear to present a fairly smoothly varying velocity structure when viewed in a display such as Fig. 6, there are clear discontinuities when examined closely.

\subsection{The properties of the narrow line width features}
Table 2 summarizes the observed properties of the clouds using the same column headings as Table 1.  For Table 2 the width of a cloud in two coordinates as measured at half peak column density is given in Col. 8.  Several of these features appear unresolved in the 9.\arcmin1 beam in one or both coordinates and in those cases, for the sake of subsequent order-of-magnitude calculations, their dimensions were set at 5\arcmin\ using clues derived from aperture synthesis data discussed in \S1. In some cases beam broadening could be measured which allowed an estimate of the angular width slightly less than the beam width.  The average width for of all these clouds to 8.\arcmin6 which, within the error bars, is the beam width.  

Table 3 lists the properties of several of the so-called transition features, or TFs, with the same format as Table 1.  Their properties do not readily match either those of the filament or the colder clouds.

Figure 7 shows area maps of column density for several clouds with respect to filamentary segments.  This is but a small selection of all the maps that were produced. Within each map frame the average center velocity and line width for the component family are indicated.  Fig. 7a overlays data for Cloud 1.1, average line width 4.0 km s$^{-1}$ (see Table 2), as an inverted gray-scale image on a contour map representation of the broad line width family, average line width 22.9 km s$^{-1}$, of components that are found in this area. Shown in Fig. 7b is a  feature in the same area but with an average line width of 9.5 km s$^{-1}$.  Its line width lies between the bulk of the values for filament segments and the clouds.  This is an example of a transition feature, labeled TF1.  Here a connection between all three types of components appears to exist, an impression reinforced by the protrusion of the broad line width component at {\it b} $=$ 25.\arcdeg45 from the bulk of the emission at the left-hand side of Fig. 7a.

Fig. 7c shows the broad line width feature labeled Cloud 5 which shows filament line width characteristics while Fig. 7d shows two associated narrow line width clouds at distinctly different velocities that appear to be slightly offset in angle from the feature in (c).  Fig. 7e shows the morphology of a transition component, TF3, that may be compared with the offset narrow line width feature, Cloud 6.1, in Fig. 6f.     

Fig. 8 shows the morphology of several broad and narrow line width components in an area where their relationships are instructive as regards key conclusions that will be drawn from these data.  Fig. 8a plots the column densities of the relatively amorphous looking broad components as an inverted gray scale image with the average center velocities and line widths for patches of emission indicated.  The average line widths of the broad component in this area are all less than the value of 23 km s$^{-1}$ referred to before in reference to the more clearly defined filament segments listed in Table 1.  This is consistent with the results of Hulsbosch (1978) referred to in \S3 who only mapped this area of A0.  The feature at the upper right is an area where both the average line width and center velocities have a large range which indicates that there the emission profile structure has contributions from a complex superposition of underlying components along the line of sight.  Similarly the amorphous structure at {\it l} $>$ 132.\arcdeg8 shows a very large range of line width values for the underlying broad components so that no clear family of values could be identified.

In contrast, this area encompasses several clouds as shown in Figs. 8b \& 8c, where they are overlain on the same gray-scale image used in Fig. 8a..  Again their average velocities and line widths are shown together with the value of the peak column densities.  The contour intervals generally range from 5 in intervals of 5 or 10 $\times$ 10$^{18}$ cm$^{-2}$.  In other cases they are less in order to introduce some uniformity of appearance in the contours.  At the top-right two narrow line width features closely spaced in velocity overly the otherwise complex broad line width structure referred to above.  Two elongated narrow line width features closely spaced in velocity are seen centered at about ({\it l,b}) $=$ (131.\arcdeg6, 23.\arcdeg5) and their properties are listed in Table 2 as PairS A and PairS B.  In both Figs. 8b \& c it appears that several of the elongated clouds are not always coincident in location with the brighter parts of the broad line width areas seen in the figure.  

\subsection{Mass distribution in the various features}
At a distance of 200 pc (discussed in the next section) the mass of gas in the filament is 0.23 $\pm$ 0.20 $M_{\sun}$ $pc^{-1}$.  The average mass of the clouds is 0.12 $\pm$ 0.11 $M_{\sun}$. The corresponding values for a distance of 4 kpc (see \S4) are 4.8 $\pm$ 4.1 $M_{\sun}$ $pc^{-1}$ in the filament and 2.3 $\pm$ 2.2 $M_{\sun}$ $pc^{-1}$ in the clouds.

\section{Properties of the HI gas}
The data displayed in Figs. 3, 7 \& 8 show that the narrow line width (cold cloud) HI features do not necessarily lie in the same direction as the broad line width (warmer) gas that defines the filament.  This contradicts a simple core-halo model to account for the two major line width components as being in pressure equilibrium since they would be nestled inside one another.  Often they are not coincident in space as inferred by their positional offset in area maps.  Also, the data in Tables 1 \& 2 show that there is no uniform kinetic temperature within the features that were mapped, if the observed line width is taken as a measure of the kinetic temperature. 

The next challenge is to convert the observed physical parameters to those that allow the physics of the filament and the clouds to be compared.  For example, what are the internal pressures?  This requires knowing a distance to convert observed column density and depths along the line-of-sight to a volume density.  Also required is the actual kinetic temperature, setting aside for a moment the reason for the large line widths in the filament segments.  

\subsection{Deriving densities}
To derive a volume density from an observed HI column density requires knowledge of the line-of-sight depth in the gas and a distance.  In the case of an HI filament another unknown comes into the picture, a problem that has also been confronted by those who study coronal loops (Schmelz et al. 2011), which are filaments on a very different linear scale. The question is whether a given loop is a single coherent structure, a flux tube, or consists of a collection of tangled strands.  The solar studies have enormously greater resolution than the present HI mapping but the same question remains.  The available HI data are not sufficient to determine if multiple strands, whose volume density would be higher than for a single flux tube, are involved and therefore our approach is to assume that the filament is a single flux tube whose average properties are determined by the Gaussian analysis.  

But what is the depth of the filament?  Is it as wide as it is deep, hence tube-like, or is its depth either greater or less than its width?   In each case the calculation of volume density is affected.  For example, if the filament was sheet-like, or in the shape of a ribbon, this would have to be taken into account.  It will be assumed that on average the depth along the line-of-sight is determined by the width across the line-of-sight because many samples of filament and cloud properties are available.  Whatever the true 3-D structure, it is unlikely that it would always be sampled where the line-of-sight is either along an elongated ribbon-like feature or normal to it.  On average, it is not unreasonable to assume that the depth along the line-of-sight is set by the width normal to it.  Then the volume densities can be derived, if the distance to A0 is known.

\subsection{Distance estimates}
This is not the forum to argue the distance to A0 so instead the theory that follows will first be applied assuming a nearby distance of about 200 pc taken from Verschuur (1993) and then it will be shown how the derived parameters for the A0 features change for a distance of 4 kpc, a value considered by, for example, van Woerden \& Wakker (2004). 

\subsection{The temperature of the HI gas}
The other crucial parameter is the kinetic temperature of the gas in each of the three types of feature revealed in the mapping data, Tables 1 $-$ 3.  In many studies of HI small-scale structure the kinetic temperature of the HI gas is taken to be of order 100 K, see for example Wakker (2001).  This may be a reasonable estimate for the temperature in the clouds but for the filament segments it is less obvious.  The observed line widths listed in Tables 1 \& 2 are all greater than the values of 2.2 km s$^{-1}$ expected for HI at 100 K.  At another extreme, if the HI in the filament were at or near 8,000 K H$\alpha$ emission would be expected.   We used the WHAM survey data of Haffner et al. (2003) to map the area of Fig. 1 and it shows no structure that can be related to the presence of A0.   Therefore as a first guess a temperature of 7,000 K will be adopted and the implications of choosing lesser values will be considered below.  Even if 7,000K was applicable, the observed average line width for filament segments of 22.8 km s$^{-1}$ cannot be accounted for by thermal motion alone since that implies a kinetic temperature of near 11,000 K, in which case the HI would not be observable at all since it would be ionized.  It is possible that turbulence has to be invoked to account for the non-thermal line widths but then it is necessary postulate that the turbulence is somehow quantized to account for the line widths of filament segments as opposed to the clouds.  This problem was already discussed by Verschuur \& Schmelz ( 2010) and in a similar vein a set of profiles toward A0 taken from the LAB survey of Kalberla et al. (2005) were also decomposed into Gaussians.  Again, an average line width of 24 km s$^{-1}$ appears to underlie the profiles.  This also suggests that beam broadening is not acting when the angular resolution is changed from 9.\arcmin1 of the GBT to the 36\arcmin\ of the LAB survey.   In \S6 an alternative explanation for the line broadening is offered invoking magnetic fields and Alfv\`en waves acting as a form of magnetic turbulence.  

\subsection{The presence of magnetic fields}
In what follows in \S5 the role of magnetic fields in determining the filament structure is described and it is worth noting in advance that there is a large body of evidence related to interstellar magnetic field structure and filaments.  This is not the forum to provide and extensive review of the evidence and only a few contributions are noted. For example, Gomez de Castro, Pudritz \& Bastien (1997) found from optical polarization data that the magnetic field associated with the anti-center stream of high velocity gas is normal to the filament axis, which would imply a toroidal component.  In contrast the HI filaments associated with the North Polar Spur exhibit field orientations normal to filament axes at high latitudes but along the axis at lower latitudes as pointed out already several decades ago in a review paper by Verschuur (1979), which summarized a great deal of data related to this issue.  A selection of other papers on the role of magnetic fields in filaments include the case of the Lynds 204 complex by McCutcheon et al. (1986), by McDavid (1984) for filaments in Cygnus, and in filamentary molecular clouds by Fiege  \& Pudritz (2000a \& b).  In contrast some have argued that hydro dynamical considerations are enough to account for filamentary structure, for example Begum et al. (2010), but our thesis is that the role of magnetic fields must be considered because of the large amount of evidence that points to the presence of magnetic fields in interstellar filamentary features.

\subsection{Average properties of the HI}
The average properties of the filament segments and the clouds can be derived from the observed properties by choosing a distance.  Then an internal pressure, n$_{H}T_{k}$, the product on the HI density and kinetic temperature, can be derived.  The results are shown in Table 4 for the average properties of the two classes of HI feature discussed above as well as several transition features (TFs) which do not obviously fit into either the cloud of filament categories.  The results are given for both 200 pc and 4 kpc distances.  In addition the pressures are calculated first using the observed line width as an indication of the kinetic temperature and then using assumed values for the kinetic temperature.  (The derived data for individual features will be shown in tables to follow.)  Table 4, Col. 1 lists the feature class and Col. 2 gives the average line width for the sample.  The average apparent kinetic temperatures derived from the observed line widths are given in Col. 3 as well as alternate values for T$_{k}$.  The average column density of the HI for a 200 pc distance is given in Col. 4 and the apparent pressure in Col. 5.  The last two columns show the values derived if a distance of 4 kpc is assumed.  The last line in the table gives the galactic halo pressure from Wolfire et al. (1995) and over the range of distances considered here that pressure does not vary significantly because at the galactic latitude of A0. 

Table 4 shows that no matter what distance is chosen or whether the observed line width indicating internal thermal motion is chosen over an alternative value for $T_{k}$, the thermal pressures of the filament segments and the clouds cannot be brought into simple pressure equilibrium with one another or with surrounding galactic halo gas.  The only way to avoid this conclusion is to invoke {\it ad hoc} assumptions about the internal temperatures and linear dimensions of each feature that would adjust the volume densities so that each has unique properties that bring it into pressure equilibrium  with its surroundings.  We do not favor this approach in the light of what is found in the analysis to follow.

Another way of looking at this problem is to recognize that the external thermal pressure required to establish localized equilibrium varies tremendously along the filament and from cloud to cloud.   For example, taking the kinetic temperature for the filament as 7,000 K and for clouds at 200 K the implied pressure difference is 16.9 times greater in the filament than in the clouds, using the averages in Table 5 \& 6, below.  Alternatively, if the observed line widths are used to indicate kinetic temperature, for which there is in fact little evidence, the range of internal pressures in the filament is 7.8 between maximum and minimum and for the clouds is it is 3.0.  These ranges prompted the exploration of the role of magnetic fields in stabilizing the filaments and clouds to determine if that would produce a more unified picture

\section{Confinement of HI in an Interstellar  Filament}
The goal of this work is to determine what controls the stability of the filament and the clouds. For many decades hydrodynamics has been invoked to describe the physics of interstellar structures.  But that is to ignore the presence of magnetic fields, and by implication electric currents.  Instead, it is suggested there that magneto-hydrodynamics, specifically plasma physical concepts, deserve to be considered.  A similar evolution of ideas long ago occurred in the study of the physics of the solar corona, which require that the role of magnetic fields has be taken into account to explain filamentary phenomena.  
 
If gas is flowing along magnetic field lines (inside flux tubes) it is not unreasonable to allow for the presence of an electric current because the motion of electrons and neutrals are coupled.  Such a current will produce a toroidal magnetic field that will act to stabilize the material in the filament against internal pressure. This hypothesis was first suggested by Bennett (1934) and the effect is known as the Bennett pinch (or Z-pinch).  Extensive discussions are to be found in Alfv\'en \& Carlqvist (1978), Carlqvist (1988), Peratt (1990, 1992, 1998), Carlqvist \& Gahm (1992), and Verschuur (1995a).  Carlqvist \& Gahm (1992) note that such pinches are inherently unstable unless the magnetic field so generated has a helical component.  In the presence of instabilities they further note that a whole spectrum of Alfv\'en waves will be generated. 

The following analysis rests on the groundwork laid by Carlqvist (1988).  It is built on the assumption that neutrals are closely coupled to the ionized component in the gas masses under consideration.  His work was in turn built on that of Bennett (1934) who showed that an axially directed current will produce a toroidal magnetic field that acts to produce an inwardly directed j x B force.  The Bennett relationship is: 
\begin{equation}
\mu_{0}/4\pi\ I^2 = 2 N k (T_{e} + T_{i}).
\end{equation}
N is the number of electrons per unit length along the axis, $T_{e}$ and $T_{i}$ are the electron and ion temperatures, I is the total beam current, and k is Boltzmann's constant.

Carlqvist (1988) considered the balance between forces in the filament acting to gather the matter, which are gravity and the toroidal magnetic field produced by the field-aligned current, {\it I} (also known as a Birkeland current).  These are counteracted by excess internal thermal pressure over any external thermal pressure due to, for example, a surrounding low density, warm or hot medium, an internal magnetic pressure, and rotation around the filament axis.  The Carlqvist relation will be evaluated using data from the high-resolution observations of A0 bearing in mind that atomic hydrogen atoms, mass $m_{H}$, volume density $n_{H}$, are coupled to the motion of electrons in the form of a current along the filament axis.  

The Carlqvist Relation (from his Eqtn 2.15) is,
\begin{equation}
\mu_{0} I^2(a) + 4\pi\ G \overline{m}^2  N^2(a) = 8 \pi[\Delta{W_{k}}(a) + \Delta{W_{Bz}}(a) + W_{rot}(a)].
\end{equation}

The two terms on the left-hand side of Eqtn. 2 are the pinching forces due to the current, {\it I}, setting up a toroidal field, $B_{\phi}$, and gravity operating in a filament of radius a. In the general form the various parameters to be discussed may vary as a function of r, the radius of a cylindrical filament, but a basic simplifying assumption will be used here to treat the current, magnetic field and density inside the filament as uniform out to radius a.  N(a) is the number of particles per unit length, $\Delta{W_{k}}(a)$ is a thermal energy term, $\Delta{W_{Bz}}(a)$ a magnetic energy term relating to the axial magnetic field, $B_{z}$, and $W_{rot}(a)$ is related to the kinetic energy of rotation, all estimated per unit length. 

In Eqtn. 2, $\overline{m}$ is defined as the  mean particle mass given by
\begin{equation}
\overline{m} = (m_{H} + m_{i} + m_{e})/n
\end{equation}
where the masses of the neutrals, ions and electrons are taken into account, and n is the total particle density given by
\begin{equation}
n = n_{H} + n_{i} + n_{e}.
\end{equation}

For small fractional ionization Eqtn. 2 can be rewritten in terms of the contribution from HI atoms, volume density $n_{H}$, temperature $T_{k}$ and setting N as the total number of hydrogen atoms per unit length in the cylindrical filament, as 
\begin{equation}
\mu_{0} I^2(a) + 4\pi\ G m_{H}^2  N^2 = 8 \pi[\Delta{W_{k}}(a) + \Delta{W_{Bz}}(a) + W_{rot}(a)].
\end{equation}

This equation is akin to the more traditional virial theorem applied to HI "clouds" (e.g. Spitzer, 1978) except that it takes into account the possible variation as a function of radius within the filament of the various key parameters, the current, temperature and axial magnetic field strength.  The relation of the various terms to observable quantities to produce Eqtn. 2 will be considered next.  

In the bulk of this paper the relevance of gravitational force is ignored because the difference between first and second terms in the left-hand side of Eqtn. 5 is 5 orders of magnitude at the distances explored here.  Even a simple calculation using the traditional virial theorem shows that the thermal terms is about three orders of magnitude larger than the gravitation energy and two energies only come into balance at distances of several Mpc. 

\subsection{The Thermal Energy Term}
The thermal energy term on the right-hand side of Eqtn. 5 consists of two contributions that quantify the relative values of internal versus external thermal energy, namely:
\begin{equation}
\Delta{W_{k}}(a)=W_k{a} - \pi a^2 p_{k}(a)
\end{equation}
where $p_{k}$ is the kinetic pressure of the particles  and the first term refers to those particles inside the filament and the second term to the pressure at the boundary of the filament of radius a.  According to the data in Table 4 for a distance of order 200 pc and using the observed line widths, the external thermal pressure is very much less than the apparent internal thermal pressure of the filament so that this second term can be ignored in initial calculations.  

Using Carlqvist's Eqtn 2.10, 
\begin{equation}
W_{k}(a) = \int _0^a 2\pi r p_{k}dr 
\end {equation}
where
\begin{equation}
p_k = n_{H} k T_{k}.
\end{equation}
Thus, integrating, 
\begin{equation}
\Delta{W_{k}}(a) = \pi a^2 n_{H} k T_{k}
\end{equation}
is the thermal pressure due to kinetic temperature of the neutral gas inside the filament.

\subsection{The Magnetic Energy Term}
The second term on the right-hand side of Eqtn. 5 is the excess magnetic energy associated with the internal magnetic field, $B_{z}$ with respect to the external magnetic field and following Carlqvist's Eqtn. 2.14, this is
\begin{equation}
\Delta{W_{Bz}}(a) = W_{Bz}(a) - \pi a^2 B_{z}^2(a)/(2\mu_{0}).
\end{equation}

The second term in Eqtn 10 relates to the external galactic magnetic field. The magnitude of this field strength is about 2.2 $\mu$G for the galactic disk (see, for example, Manchester 1974).   Very little is known as regards the magnetic field strength in the galactic halo but it is almost surely $<$1 $\mu$G.  These values are less than the internal magnetic field strength if a 200 pc distance is used, which will become apparent when these quantities are evaluated.  This will be discussed further in \S8. 

The first term in the RHS of Eqtn. 10 is the internal magnetic energy given by Carlqvist's Eqtn. 2.11 which is
\begin{equation}
W_{Bz}(a) = \int _0^a 2\pi r  B_{z}^2/(2 \mu_{0}) dr.
\end{equation}
Assuming a uniform field inside the filament as a function of radius, integrating, and substituting in Eqtn. 10 gives
\begin{equation}
\Delta{W_{Bz}}(a) = \pi a^2 B_{z}^2/(2 \mu_{0}).
\end{equation}

The strength of the axial magnetic field, $B_{z}$, would appear to be unknown but the essence of the argument to follow in \S6 below is that it may be derived from the observed line width by invoking the presence of Alfv\`en waves.  

\subsection{The Rotational energy Term}
The third term on the right-hand side of Eqtn. 5 is the rotational energy which must be included in the case of the segment of the filament called ARC, which shows a very large velocity of rotation.  From Eqtn. 2.12 of Carlqvist
\begin{equation}
W_{rot}(a) = \int _0^a  \pi r m_{H} n_{H}  \omega^2 {r}^2 dr.
\end{equation}
The angular velocity, $\omega$ is related to the observed parameter $v_{a}$, the velocity of rotation measured at the edge of the filament at radius a. Thus replacing  $\omega$ by $v_{a}/a$ and integrating we obtain
\begin{equation}
W_{rot}(a) = \pi a^2 m_{H} n_{H} v_{a}^2.
\end{equation}

\subsection{The Toroidal Magnetic Field}
Eqtn. 5 can now be rewritten in terms of observable quantities and recognizing that the gravity term plays no role in the energy balance.
Thus, using Eqtns. 9, 12 \& 14,
\begin{equation}
\mu_{0} I^2(a)  = 8 \pi[\pi a^2 n_{H} k T_{k}) + \pi a^2 B_{z}^2/(2 \mu_{0}) + \pi a^2 m_{H} n_{H} v_{a}^2].
\end{equation}
Hence
\begin{equation}
\mu_{0} I^2(a)  = 8 \pi^2 a^2[ n_{H} k T_{k}) +  B_{z}^2/(2 \mu_{0}) +  m_{H} n_{H} v_{a}^2].
\end{equation}

The key ingredient to confining the gas to the filament is then the toroidal magnetic field, $B_{\phi}$, produced by the current, $I_{a}$.  In order to relate this to observable quantities consider that a toroidal magnetic field produced by a current is given by
\begin{equation}
 B_{\phi}(a) = \mu_{0} I(a)/(2 \pi a). 
  \end{equation}
 Hence 
 \begin{equation}
  I(a) = 2 \pi a  B_{\phi}(a)/\mu_{0}.
  \end{equation}
 
 Combining Eqtns. 18 \& 16 and setting $\mu_{0}$ = 1, gives:
 \begin{equation}
  B_{\phi}(a) = 2^{1/2}[ n_{H} k T_{k}) +  B_{z}^2/2 +  m_{H} n_{H} v_{a}^2]^{1/2}
  \end{equation}

If the axial magnetic field strength is known, this equation can be applied to the HI parameters for the features mapped and listed in Tables 1, 2 \& 3.

\section{Excess line width, Alfv\'en waves, and the axial magnetic field strength}
The results of Gaussian analysis of the HI profiles observed toward high-velocity feature A0 shows that the line widths are very much larger than expected for reasonable estimates of the kinetic temperature of neutral atomic hydrogen gas.  It is hypothesized that the observed line widths are due to broadening by the presence of Alfv\'en waves in the gas encompassed by the telescope beam.  An ensemble of Alfv\'en waves with velocity V$_{A}$ within the telescope beam will act to broaden the HI emission profile and contribute to the observed line width, ${\bf W}_{obs}$.  Alfv\'en waves are transverse waves that propagate along magnetic field lines. V$_{A}$ represents the velocity of the wave along the magnetic field axis and may not necessarily equate to the transverse motion experienced by the hydrogen atoms as the wave passes.  For an order of magnitude calculation it is assumed that the transverse motion can reach a maximum velocity of V$_{A}$, although it depends on the amplitude of the wave, which is unknown.  However, transverse waves do not contribute to pressure in the gas.  Also, bearing in mind the point made by Carlqvist \& Gahm (1992), that in the presence of instabilities a whole spectrum of Alfv\`en waves will be generated, we can consider the ensemble to act as a form of turbulence that determines the observed line width in excess of purely thermal motions. 

In the presence of Alfv\'en waves in the beam, the observed line width, {\bf W}$_{obs}$ is given by, 
\begin{equation}
	{\bf W}_{obs}^{2} = {\bf W}_{th}^{2}  +  {\bf W}_{AW}^{2},	
\end{equation}
where {\bf W}$_{th}$ is the thermal line width and {\bf W}$_{AW}$ is the line width due to Alfv\'en waves.  If the kinetic temperature is known, or if reasonable values can be considered (see \S 4 above), calculation of the possible line broadening due to Alfv\'en waves is possible. 

The Alfv\'en velocity, V$_{A}$, is given by 
\begin{equation}
V_{A}^2 = B_{AW}^2/4\pi\rho,    						
\end{equation}
so that
\begin{equation}
B_{AW}=(4\pi\rho)^{1/2}\  V_{A} =  B_{z},    						
\end{equation}
where $B_{AW}$ is the magnetic field strength derived by using the Alfv\'en velocity and $\rho$ is the mass density of the HI gas.  The value of V$_{A}$ is obtained from the dispersion of the line width {\bf W}$_{AW}$ due to the Alfv\'en waves using Eqtn. 20.  This value of the magnetic field is set to the value of the axial magnetic field, $B_{z}$.  Thus Eqtn. 19 can be rewritten, again after treating the field within the filament as independent of radius, as
\begin{equation}
  B_{\phi} = 2^{1/2}[ n_{H} k T_{k}) +  2\pi\rho\  V_{A}^2 +  m_{H} n_{H} v_{a}^2]^{1/2}
  \end{equation}

Because the Alfv\'en velocity does not contribute to the energy balance in the filament it will allow the strength of the magnetic field inside the filament to be derived from the observed line widths for a given kinetic temperature.  

\section{Application of the Filament Theorem}
Eqtn. 23 can be evaluated for each of the classes of features discussed above.  As a starting point a distance of 200 pc is chosen and the implications of increasing this to 4 kpc pc will be illustrated in \S8 below. In what follows the kinetic temperatures are also required and the chosen values will be discussed in context. 

\subsection{Derived Properties of the Filament Segments}
Table 5 shows the values of the derived properties for the filament segments using the observed properties listed in Table 1. Here a kinetic temperature of 7,000 K is used, a value discussed in \S4 and the consequence of using different values will be discussed in \S8 below.  Col. 1 gives the name of the filament segments from Table 1 and Col. 2 is the depth along the line-of-sight assuming the segment is as deep as it is wide.  Col. 3 gives the derived volume density along the line-of-sight for the distance of 200 pc.  Col. 4 gives the pressure, the product of the kinetic temperature indicated in the table head and the volume density. The contribution to the observed line width, ${\bf W}_{AW} $, due to the Alfv\'en waves as per \S6, is given in Col. 5. This is the full width at half maximum of the Gaussian due to an ensemble of those waves inside the beam.  The magnetic field that gives rise to these Alfv\'en waves is given in Col. 6 using Eqtn. 22.  Col. 7 gives the value of the derived field-aligned current and Col. 8 gives the value of the corresponding toroidal magnetic field. Also shown in Table 5 are the average values and standard deviations for the various parameters excluding ARC, which is listed separately because it manifests a large rotational velocity and the contribution from its rotational energy has been included in the full evaluation of Eqtn. 23, as shown in Cols. 7 \& 8.  It is striking that inclusion of rotation in the case of ARC brings the strength of required toroidal magnetic field, 5.8 $\mu$G, into line with that for the average in the other segments in contrast to the value caculated without rotation, which is 3.8 $\mu$G.  It is also striking that the average axial magnetic field strength for the filament segments of 5.6 $\mu$G is comparable to the value of 6.5 $\mu$G for the toroidal field because of the relative unimportance of the internal thermal energy terms in comparison with the magnetic energy term. 

The average derived current value of 9.7 x 10$^{13}$ Amps for the 200 pc distance in Table 4 is close to the value estimated for unrelated and nearby filamentary features discussed by Carlqvist (1988) of 4 $-$ 7 x 10$^{13}$ Amps while the value obtained by Verschuur (1995a) for nearby filaments of 4 - 11 x 10$^{13}$ Amps is essentially the same.  If, however, the distance of the filament segments listed in Table 5 is 4 kpc the average derived current would be 43.4 x 10$^{13}$ Amps, a factor of 5 greater than for filaments known to be nearby. If future work on the distance to the HI in this study was to unambiguously confirm a value in the kpc range then the above analysis would have to include the role of external thermal pressure as well as the external magnetic field strength, because the value of the derived field strengths for the 4 kpc distance are of the same order as the external field of about 1 $\mu$G, see discussion in \S9 below. 

\subsection{Derived Properties of the clouds}
Table 6 gives the derived properties of the narrow line width clouds for a distance of 200 pc and a temperature of 100 K, assuming that they, too, are magnetically confined.    The implications of choosing a much larger distance of 4 kpc will be shown in \S8.  The columns are the same as for Table 5.  The average current for the clouds at a distance of 200 pc is 0.36 x 10$^{13}$ Amps, a factor of 2.5 times lower than for the filament segments.  This difference almost certainly contains clues to the physics of how the clouds are produced by, for example, extreme pinching of the filament.  However, the average axial magnetic field strengths obtained from the Alfv\`en velocity argument above are very similar, 5.6 $\mu$G for the filament segments compared with 4.3 $\mu$G for the clouds, while the average toroidal fields for the two classes of feature are 6.5 \& 4.4 $\mu$G respectively.  Table 7 is similar to Table 6 but for the transition features.  The derived current and magnetic field values are surprisingly similar to those for the clouds and the filament segments, but falling between the two extremes.

\subsection{The Dependence on Distance and Temperature}
The data summarized in Tables 5$-$ 7 assumed that the distance to the high-velocity feature A0 is 200 pc. In Fig. 9 the dependence on distance and temperature of the values of the axial magnetic fields derived from the Alfv\`en velocity, $B_{AW}$, as discussed in \S6, and the value of the average toroidal magnetic field, $B_{\phi}$, are shown.  Fig. 9a shows the data for the filament fields with the solid line indicating the average toroidal field and the dashed line the axial field strength for 200 pc, top two lines, and 4,000 pc, lower two lines.  Fig. 9b plots the same for the average fields in the clouds.  

The distance dependence shown in Fig. 9 is not great and the strength of the derived toroidal magnetic field lies in the range of a few ${\mu}$G to at most 10 ${\mu}$G.  The different values of the magnetic field strength in the filament for the two different distances offers a potential way to determine which is correct.  That requires direct detection of the magnetic field signature in, for example, filament segments.  There exists a reliable upper limit to the line-of-sight component of a possible field in A0 measured using the Zeeman effect technique toward the central bright feature in A0 at {\it l,b} $=$ 131.\arcdeg7, 23.\arcdeg44.  The observations were made with the 140-ft radio telescope of the NRAO by Verschuur (1995b) who found no evidence for a magnetic field detection with the observed value being $-$0.1$\pm$1.8 $\mu$G where the error is one standard deviation. That author explained why his results were in such stark contrast to a claimed field detection of  $-$11.4 $\pm$ 2.4 $\mu$G in the same direction by Kaz\'es et al. (1991).  The Verschuur (1995b) upper limit is of order 3.6 $\mu$G at the two standard deviation level, not enough to test the hypothesis suggested above.  In addition, having both an axial field and a toroidal field orientation encompassed in the large beam width used in the Zeeman effect measurments makes it highly unlikely that, because of field reversals in the beam, anything would be detected.  Perhaps some brave individuals may yet pursue this challenge in the future. An intriguing alternative would be to determine the rotation measures of distant radio sources that might happen to lie in the direction of the one side or the other of a filament segment since the toroidal magnetic field would show a reversal across the filament width.

\subsection{On the efficacy of using magnetic fields}
Examination of the data in Tables 5 \& 6 in particular show that the derived magnetic field strengths, either $B_{AW}$ or $B_{\phi}$, vary only slightly between the filament segments and the clouds, in stark contrast to the large range of thermal pressures required to balance the internal energies of the two classes of object, see \S4.5.  Not only are the ranges within each class (see Table 5 \& 6) quite small, but the average values of both magnetic field types are nearly identical in the filament and the clouds.  Thus it appears that axial and toroidal magnetic fields instead of external pressure may be the key to determining equilibrium conditions in the A0 filament and clouds.

\section{Conclusions}
Gaussian analysis of HI emission line profiles using high-resolution, pencil beam observations of the HI structure in high-velocity feature A0 show that about 70\% of the gas is in the form of a twisted filament with intrinsic line width of order 24 km s$^{-1}$ with the remaining 30\% in the form of bright, narrow line width features or clouds. A third class of features containing a small fraction of the total HI column densities but with line widths between those of the filament segments and clouds hints at the possible existence of a transition class between the former two.  Because the clouds are not always spatially coincident with the filament gas, and because the internal thermal pressures of the two types of feature are not equal, no matter what distance is chosen, it seems very unlikely that a simple core-halo model can be successfully invoked to account for the stability of the HI structures. 

A model is proposed that accounts for the confinement of HI gas to a filamentary feature by invoking a toroidal field of about 6 $\mu$G produced by the flow of gas (a current) along the filamentary axis.  The current is guided by an axial magnetic field whose magnitude is estimated by recognizing that the observed HI emission line widths may be determined by the presence of Alfv\`en waves in addition to true thermal line widths.  A spectrum of Alfv\`en waves can give rise to the appearance of turbulence being the primary determinant of observed line widths.  Research on the physics of this model, which represents a manifestation of the so-called Z-pinch or Bennett pinch, which is in turn related to a field aligned current, sometimes referred to as Birkeland current, has, to date, been confined almost exclusively to the plasma physics literature.  What is offered here is a first-order attempt to relate the theory to data obtained by radio astronomical mapping of an interstellar HI feature with a view to stimulating discussion of the proposed mechanism(s) in astrophysical situations.  Carlqvist \& Gahm (1992) did apply the theory to the physics of molecular clouds but that work triggered very little response in the astrophysical community.  Similarly, Verschuur (1995a) applied the Carlqvist (1988) model to observation of interstellar HI filaments but the data used then were not of the quality obtained in the mapping of the high-velocity feature A0 using the Green Bank Telescope of the NRAO with its 9.\arcmin1 beam.

A fuller understanding of the nature of the anomalous velocity HI features would ideally require observations with a pencil beam with angular resolution of order 1\arcmin\ so that the full picture can be derived.  This same point was made by Hulsbosch (1978), but that goal may never be realized.  Instead we must rely on aperture synthesis observations for the small-scale details, which unfortunately lose information regarding the extended emission from the filaments.  Available pencil beams can map the filaments  but cannot see details of the clouds.

I am grateful to D.L. Nidever for setting up the observing protocols and providing the data in a form I could use and for feedback on earlier versions of this paper, to J.T. Schmelz for helpful discussions and encouragement during the entire project, and Z. Kinnare is thanked for help in running the Gauss fit program during the earliest phases of the project.

{}

\begin{deluxetable}{cccccccc}															
\tablecolumns{8}					
\tablewidth{0pc}				
\tablecaption{Observed Properties of Filament Segments}								
\tablehead{																		\colhead{Name} & \colhead{{\it l}}   & \colhead{{\it b}}    & \colhead{Ave $V_c$}  & \colhead{Ave ${\bf W}_{obs}$}    & \colhead{App $T_k$}   & \colhead{Peak $N_H$}    & \colhead{Width}   \\{} & {(\arcdeg) }  &{(\arcdeg)} & {(km s$^{-1}$)} & {(km s$^{-1}$)} & {(K)} & {(10$^{18}$ cm$^{-2}$)} &{(\arcmin)} }			\startdata		
(1) & (2) & (3) & (4) & (5) & (6) & (7)&(8)\\																	
Cswy C	&	133.9	&	27.1	&	 -170.6$\pm$2.8	&	19.3$\pm$3.2	&	7822	&	11.2	&	14.3		\\
Cswy B	&	134.0	&	26.9	&	 -148.5$\pm$2.3	&	20.7$\pm$8.7	&	8998	&	10.4	&	14.3		\\
{}	&	{}	&	{}	&	{}	&	{}	&	{}	&	{}	&	{}	\\
UMB (2)	&	134.2	&	23.6	&	 -147.6$\pm$3.7	&	23.3$\pm$5.0	&	11401	&	26.2	&	33.8		\\
{}	&	{}	&	{}	&	{}	&	{}	&	{}	&	{}	&	{}		\\
Bar 1 (2)	&	135.1	&	26.1	&	 -157.0$\pm$4.6	&	23.9$\pm$4.3	&	11995	&	34.8	&	27.6	\\
Bar 2 (2)	&	134.6	&	26.1	&	 -139.7$\pm$5.2	&	26.9$\pm$7.0	&	15196	&	18.5	&	30.7		\\
{}	&	{}	&	{}	&	{}	&	{}	&	{}	&	{}	&	{}		\\
OddArea	&	135.8	&	25.1	&	 -160.7$\pm$2.4	&	24.0$\pm$3.9	&	12096	&	20.1	&	23.6		\\
{}	&	{}	&	{}	&	{}	&	{}	&	{}	&	{}	&	{}		\\
Strip 2	&	131.3	&	23.1	&	 -206.5$\pm$1.4	&	21.0$\pm$3.4	&	9261	&	57.1	&	35.0		\\
{}	&	{}	&	{}	&	{}	&	{}	&	{}	&	{}	&	{}		\\
Cloud 5	&	135.8	&	27.2	&	 -165.4$\pm$4.4	&	21.4$\pm$3.6	&	9617	&	39.5	&	31.9		\\
{}	&	{}	&	{}	&	{}	&	{}	&	{}	&	{}	&	{}		\\
Outlier (2)	&	133.4	&	27.2	&	 -171.6$\pm$3.6	&	21.4$\pm$3.2	&	9617	&	18.5	&	20.7	\\
{}	&	{}	&	{}	&	{}	&	{}	&	{}	&	{}	&	{}		\\
ARC	&	136.3	&	25.5	&	-125	&	25.0$\pm$2.0	&	10913	&	5.3	&	30.0		\\
{}	&	{}	&	{}	&	{}	&	{}	&	{}	&	{}	&	{}	\\
{}	&	{}	&	{}	&	Average	&	22.8$\pm$1.1	&	10692$\pm$1999	&	24.2$\pm$14.9	&	26.2$\pm$7.2			\\
\enddata
\end{deluxetable}
\clearpage

\begin{deluxetable}{cccccccc}															
\tablecolumns{8}																	
\tablewidth{0pc}																	
\tablecaption{Observed Properties of the Clouds}							
\tablehead{																		\colhead{Name} & \colhead{{\it l}}   & \colhead{{\it b}}    & \colhead{Ave $V_c$}  & \colhead{Ave ${\bf W}_{obs}$}    & \colhead{App $T_k$}   & \colhead{Peak $N_H$}    & \colhead{Width} \\{} & {(\arcdeg) }  &{(\arcdeg)} & {(km s$^{-1}$)} & {(km s$^{-1}$)} & {(K)} & {(10$^{18}$ cm$^{-2}$)} &{(\arcmin)}  }			\startdata			
(1) & (2) & (3) & (4) & (5) & (6) & (7)&(8)\\													
1.1	&	132.7	&	25.35	&	 -205.6$\pm$1.7	&	4.0$\pm$1.3	&	336	&	20.1	&	9.4x5.0		\\
{}	&	{}	&	{}	&	 {}	&	{}	&	{}	&	{}	&	{}		\\
2.1	&	132.4	&	24.85	&	 -202.1$\pm$1.3	&	4.7$\pm$2.5	&	455	&	15.5	&	3.8x8.1		\\
2.2	&	132.4	&	24.85	&	 -199.2$\pm$2.4	&	6.4$\pm$2.6	&	852	&	11.5	&	6.0x7.2		\\
{}	&	{}	&	{}	&	 {}	&	{}	&	{}	&	{}	&	{}		\\
3.1	&	137.4	&	27.85	&	 -210.2$\pm$0.4	&	4.5$\pm$1.1	&	425	&	37.1	&	10.0x8.6		\\
3.2	&	137.4	&	27.75	&	 -201.0$\pm$1.4	&	6.9$\pm$1.1	&	1000	&	11.0	&	7.9x8.9		\\
{}	&	{}	&	{}	&	 {}	&	{}	&	{}	&	{}	&	{}		\\
4.1	&	132.7	&	23.15	&	 -187.8$\pm$1.6	&	5.3$\pm$1.6	&	590	&	55.3	&	16.1	\\
{}	&	{}	&	{}	&	 {}	&	{}	&	{}	&	{}	&	{}		\\
5.1	&	135.6	&	27.0	&	 -177.8$\pm$1.0	&	6.6$\pm$1.5	&	915	&	13.0	&	7.6x5.0		\\
5.2	&	135.7	&	27.2	&	 -167.1$\pm$0.8	&	5.8$\pm$1.4	&	706	&	20.5	&	4.7x7.5		\\
{}	&	{}	&	{}	&	 {}	&	{}	&	{}	&	{}	&	{}		\\
6.1	&	131.8	&	26.6	&	 -198.2$\pm$1.6	&	7.6$\pm$1.9	&	1213	&	17.1	&	5.7x11.2		\\
{}	&	{}	&	{}	&	 {}	&	{}	&	{}	&	{}	&	{}		\\
PairN F	&	131.2	&	23.9	&	 -215.2$\pm$1.2	&	4.4$\pm$2.5	&	407	&	74.5	&	7.4		\\
PairN G	&	131.2	&	23.9	&	 -217.2$\pm$1.6	&	4.2$\pm$2.1	&	370	&	35.0	&	9.5		\\
{}	&	{}	&	{}	&	 {}	&	{}	&	{}	&	{}	&	{}		\\
PairS A	&	131.7	&	22.5	&	 -209.7$\pm$2.1	&	4.3$\pm$1.2	&	388	&	87.3	&	10.9		\\
PairS B	&	131.6	&	23.45	&	 -211.3$\pm$1.9	&	5.3$\pm$1.9	&	590	&	70.4	&	12.3		\\
{}	&	{}	&	{}	&	 {}	&	{}	&	{}	&	{}	&	{}		\\
Bar C.2	&	134.8	&	26.2	&	 -167.4$\pm$4.7	&	6.9$\pm$2.0	&	1000	&	14.9	&	20.0x10.1		\\
{}	&	{}	&	{}	&	 {}	&	{}	&	{}	&	{}	&	{}		\\
Outlier	&	133.0	&	27.3	&	 -181.3$\pm$4.2	&	7.7$\pm$2.8	&	1245	&	32.2	&	15.3x15.2		\\
{}	&	{}	&	{}	&	 {}	&	{}	&	{}	&	{}	&	{}		\\
{}	&	{}	&	{}	&	Average	&	5.5$\pm$0.4	&	699$\pm$315	&	34.5$\pm$25.5	&	{}		\\
\enddata																			
\end{deluxetable}																	

\clearpage

\begin{deluxetable}{cccccccc}															
\tablecolumns{8}					
\tablewidth{0pc}				
\tablecaption{Observed Properties of the "Transition" Features}								
\tablehead{			
\colhead{Name} & \colhead{{\it l}}   & \colhead{{\it b}}    & \colhead{Ave $V_c$}  & \colhead{Ave ${\bf W}_{obs}$}    & \colhead{App $T_k$}   & \colhead{Peak $N_H$}    & \colhead{Width}  \\{} & {(\arcdeg) }  &{(\arcdeg)} & {(km s$^{-1}$)} & {(km s$^{-1}$)} & {(K)} & {(10$^{18}$ cm$^{-2}$)} &{(\arcmin)}  }					\startdata																			
(1) & (2) & (3) & (4) & (5) & (6) & (7)&(8)\\													
Cloud 1.2 (2)	&	132.7	&	25.4	&	 -202.5$\pm$3.7	&	9.5$\pm$2.9	&	1895		&	41.6	&	12.0	\\
{}	&	{}	&	{}	&	{}	&	{}	&	{}	&	{}	&	{}	\\
PairN (2)	&	131.2	&	23.9	&	 -207.1$\pm$4.9	&	16.1$\pm$4.9	&	5443		&	48.9	&	27.3	\\
{}	&	{}	&	{}	&	{}	&	{}	&	{}	&	{}	&	{}	\\
Cloud 6.2 (2)	&	132.0	&	26.7	&	 -189.2$\pm$5.3	&	15.8$\pm$2.7	&	5242		&	12.6		&	17.9	\\
{}	&	{}	&	{}	&	 {}	&	{}	&	{}	&	{}	&	{}	\\
Bar C.1 (2)	&	135.2	&	26.0	&	 -163.3$\pm$2.4	&	9.1$\pm$3.3	&	1739	&	18.2	&	10.1	\\
{}	&	{}	&	{}	&	{}	&	{}	&	{}	&	{}	&	{}	\\
{}	&	{}	&	{}	&	Average	&	11.7$\pm$1.4	&	3580$\pm$2038	&	30.3$\pm$17.6	&	16.8$\pm$7.8			\\
\enddata
\end{deluxetable}

\clearpage

\begin{deluxetable}{ccccccc}													
\tablecolumns{7}													
													
\tablecaption{Average Properties of HI features in A0}													
\tablehead{													
													
\colhead{} & {} & Ave. & {At 200 pc} & {Apparent} & {At 4000 pc} & {Apparent}\\{Feature} & {Ave ${\bf W}_{obs}$}    & \colhead{App. $T_k$}   & \colhead{Ave. $n_H$}     & \colhead{Pressure} & \colhead{Ave $n_H$} & \colhead{Pressure} \\{class} &  {(km s$^{-1}$)}  & {(K)} & {(cm$^{-3}$)}  & {(cm$^{-3}$ K)} & {(cm$^{-3}$)}  & {(cm$^{-3}$ K)}}													
\startdata													
(1)	&	(2)	&	(3)	&	(4)	&	(5)	&	(6)	&	(7)	\\
Filament data	&	22.8	&	10,917	&	5.55	&	60,589	&	0.28	&	3,056	\\
Alternate $T_{k}$	&	-	&	7,000	&	"	&	38,850	&	"	&	1,960	\\
Alternate $T_{k}$	&	-	&	100	&	"	&	555	&	"	&	28	\\
{}	&	{}	&	{}	&	{}	&	{}	&	{}	&	{}	\\
TFs	&	11.7	&	2,874	&	11.2	&	32,119	&	0.56	&	2,005	\\
Alternate $T_{k}$ &  -  &  1000 &  "  & 11,200 &  "  & 560\\
Alternate $T_{k}$ &  -  &  100 &  "  & 1,120 &  "  & 56\\
{}	&	{}	&	{}	&	{}	&	{}	&	{}	&	{}	\\
Cloud data	&	5.5	&	699	&	23.0	&	16,077	&	1.13	&	665	\\
Alternate $T_{k}$	&	-	&	100	&	"	&	2,300	&	"	&	113	\\
{}	&	{}	&	{}	&	{}	&	{}	&	{}	&	{}	\\
Galactic halo pressure	&	-	&	-	&	-	&	4,000	&	-	&	4,000	\\

\enddata													
\end{deluxetable}													
\clearpage			

\begin{deluxetable}{cccccccc}														
\tablecolumns{8}														
\tablewidth{0pc}														
\tablecaption{Derived Properties of Filament Segments at 200 pc (7000 K)}														
\tablehead{														
\colhead{Name} & \colhead{Depth} & \colhead{$n_H$} &\colhead{$n_H$T}  & \colhead{${\bf W}_{AW}$} & \colhead{$B_{AW}$} & \colhead{Current} & \colhead{$B_{tor}$} \\ {}  &{(pc)} & {(cm$^{-3}$)} &   {(cm$^{-3}$ K)}     & {(km s$^{-1}$)}& {($\mu$G)} & {(10$^{13}$ A)} & {($\mu$G)} } 																\startdata														
(1) & (2) & (3) & (4) & (5) & (6) & (7) & (8)\\														
Cswy C	&	0.84	&	4.45	&	31167	&	6.3	&	2.5	& 	3.1	&	3.9\\
Cswy B	&	0.84	&	4.16	&	29089	&	9.8	&	3.8	& 	3.8	&	4.8\\
{}	&	{}	&	{}	&	{}	&	{}	&	{}	&	{}	&	 {}\\
UMB 	&	1.98	&	4.41	&	30893	&	14.5	&	5.9	& 	12.2	&	6.5\\
{}	&	{}	&	{}	&	{}	&	{}	&	{}	&	{}	&	 {}\\
Bar 1 	&	1.61	&	7.18	&	50290	&	15.4	&	8.0	&	13.4	&	8.8\\
Bar 2 	&	1.80	&	3.44	&	24058	&	19.8	&	7.0	&	12.7	&	7.5\\
{}	&	{}	&	{}	&	{}	&	{}	&	{}	&	{}	&	 {}\\
Odd Area	&	1.38	&	4.86	&	33993	&	15.6	&	6.6	& 	9.5	&	7.3\\
{}	&	{}	&	{}	&	{}	&	{}	&	{}	&	{}	&	 {}\\
Strip 2	&	2.05	&	9.3	&	65126	&	10.4	&	6.1	 &	14.3	&	7.4\\
{}	&	{}	&	{}	&	{}	&	{}	&	{}	&	{}	&	 {}\\
Cloud 5	&	1.87	&	7.05	&	41366	&	11.2	&	5.7	&	11.9	&	6.8\\
{}	&	{}	&	{}	&	{}	&	{}	&	{}	&	{}	&	 {}\\
Outlier	&	1.21	&	5.1	&	35600	&	11.2	&	4.8	&	6.6	&	5.8\\
{}	&	{}	&	{}	&	{}	&	{}	&	{}	&	{}	&	 {}\\
Average	&	1.51	&	5.55	&	38871	&	12.7	&	5.6	&	9.7	&	6.5\\
{}	&	$\pm$0.47	&	$\pm$1.89	&	$\pm$13238	&	$\pm$4.0	&	$\pm$1.7	&	$\pm$4.2	&	$\pm$1.5\\
{}	&	{}	&	{}	&	{}	&	{}	&	{}	&	{}	&	 {}\\
ARC  	&  	1.75	&	1.01	&	7074	&	17.1	&	3.3	&	9.7	&	5.8\\
{}	&	{}	&	{}	&	{}	&	{}	&	{}	&	{}	&	 {}\\
\enddata														
\end{deluxetable}

\begin{deluxetable}{cccccccc}																		
\tablecolumns{8}																				
\tablewidth{0pc}																				
\tablecaption{Derived Properties of the Clouds at 200 pc (100 K)}					
\tablehead{																				
\colhead{Name} & \colhead{Depth} & \colhead{$n_H$} &\colhead{$n_H$T}  & \colhead{${\bf W}_{AW}$} & \colhead{$B_{AW}$} & \colhead{Current} & \colhead{$B_{tor}$} \\ {}  &{(pc)} & {(cm$^{-3}$)} &   {(cm$^{-3}$ K)}     & {(km s$^{-1}$)}& {($\mu$G)} & {(10$^{13}$ A)} & {($\mu$G)} } 															
\startdata																				
(1) & (2) & (3) & (4) & (5) & (6) & (7) & (8)\\													
1.1	&	0.29	&	22.92	&	2292	&	3.4	&	3.1	&	0.22	&	3.2	\\
{}	&	{}	&	{}	&	{}	&	{}	&	{}	&	{}	&	{}	\\
2.1	&	0.22	&	23.30	&	2330	&	4.1	&	3.8	&	0.17	&	3.9	\\
2.2	&	0.35	&	10.91	&	1091	&	6.0	&	3.8	&	0.18	&	3.8	\\
{}	&	{}	&	{}	&	{}	&	{}	&	{}	&	{}	&	{}	\\
3.1	&	0.50	&	24.61	&	2461	&	3.9	&	3.8	&	0.39	&	3.8	\\
3.2	&	0.46	&	7.91	&	791	&	6.5	&	3.5	&	0.20	&	3.6	\\
{}	&	{}	&	{}	&	{}	&	{}	&	{}	&	{}	&	{}	\\
4.1	&	0.94	&	19.59	&	1959	&	4.8	&	4.1	&	0.65	&	4.2	\\
{}	&	{}	&	{}	&	{}	&	{}	&	{}	&	{}	&	{}	\\
5.1	&	0.29	&	14.80	&	1480	&	6.2	&	4.6	&	0.18	&	4.7	\\
5.2	&	0.27	&	24.90	&	2490	&	5.4	&	5.2	&	0.21	&	5.2	\\
{}	&	{}	&	{}	&	{}	&	{}	&	{}	&	{}	&	{}	\\
6	&	0.33	&	17.13	&	1713	&	7.3	&	5.8	&	0.22	&	5.8	\\
{}	&	{}	&	{}	&	{}	&	{}	&	{}	&	{}	&	{}	\\
PairN F	&	0.43	&	57.36	&	5736	&	3.8	&	5.6	&	0.51	&	5.7	\\
PairN G	&	0.56	&	21.00	&	2100	&	3.6	&	3.2	&	0.40	&	3.3	\\
{}	&	{}	&	{}	&	{}	&	{}	&	{}	&	{}	&	{}	\\
PairS A	&	0.64	&	45.67	&	4567	&	3.7	&	4.8	&	0.68	&	5.0	\\
PairS B	&	0.72	&	32.64	&	3264	&	4.8	&	5.3	&	0.64	&	5.4	\\
{}	&	{}	&	{}	&	{}	&	{}	&	{}	&	{}	&	{}	\\
Bar C.2	&	0.59	&	8.41	&	841	&	6.5	&	3.7	&	0.27	&	3.7	\\
{}	&	{}	&	{}	&	{}	&	{}	&	{}	&	{}	&	{}	\\
Outlier	&	0.89	&	12.09	&	1209	&	7.4	&	4.9	&	0.48	&	5.0	\\
{}	&	{}	&	{}	&	{}	&	{}	&	{}	&	{}	&	{}	\\
Average	&	0.50	&	22.88	&	2288	&	5.0	&	4.3	&	0.36	&	4.4	\\
{}	&	$\pm$0.22	&	$\pm$13.67	&	$\pm$1367	&	$\pm$1.4	&	$\pm$0.9	&	$\pm$0.19	&	$\pm$0.9	\\
{}	&	{}	&	{}	&	{}	&	{}	&	{}	&	{}	&	{}	\\
\enddata							
\end{deluxetable}	

\clearpage

\begin{deluxetable}{ccccccccc}															
\tablecolumns{8}													
\tablewidth{0pc}													
\tablecaption{Derived Properties of "Transition" Features at 200 pc (1000 K)}							
\tablehead{													
\colhead{Name} & \colhead{Depth} & \colhead{$n_H$} &\colhead{$n_H$T}  & \colhead{${\bf W}_{AW}$} & \colhead{$B_{AW}$} & \colhead{Current} & \colhead{$B_{tor}$}  \\ {}  &{(pc)} & {(cm$^{-3}$)} &   {(cm$^{-3}$ K)}     & {(km s$^{-1}$)}& {($\mu$G)} & {(10$^{13}$ A)}  & {($\mu$G)}} 	
\startdata												
(1) & (2) & (3) & (4) & (5) & (6) & (7) & (8)\\												
Cloud 1.2 	&	0.70	&	19.76	&	19758	&	6.5	&	5.6	& 	1.5	&	6.1	\\
{}	&	{}	&	{}	&	{}	&	{}	&	{}	&	{}		& 	{}\\
Pair N	&	1.60	&	10.22	&	10217	&	14.5	&	8.9	& 	2.5		& 9.1	\\
{}	&	{}	&	{}	&	{}	&	{}	&	{}	&	{}		& 	{}\\
Cloud 6.2	&	1.05	&	4.00	&	3997	&	14.2	&	5.5	&	1.0		&	5.6	\\
{}	&	{}	&	{}	&	{}	&	{}	&	{}	&	{}		& 	{}\\
Bar C1	&	0.59	&	10.93	&	10927	&	5.9	&	3.8	& 	1.0		&	4.2	\\
{}	&	{}	&	{}	&	{}	&	{}	&	{}	&	{}		& 	{}\\
Average	&	0.98	&	11.22	&	11225	&	10.3	&	5.9	&	1.5		&	6.2	\\
{}	&	$\pm$0.45	&	$\pm$6.48	&	$\pm$6485	&	$\pm$4.7	&	$\pm$2.2	&	$\pm$0.7		&$\pm$2.1 \\
{}	&	{}	&	{}	&	{}	&	{}	&	{}	&	{}		& 	{}\\

\enddata													
\end{deluxetable}

 \begin{figure}
\figurenum{1}
\epsscale{1.0}
\plotone{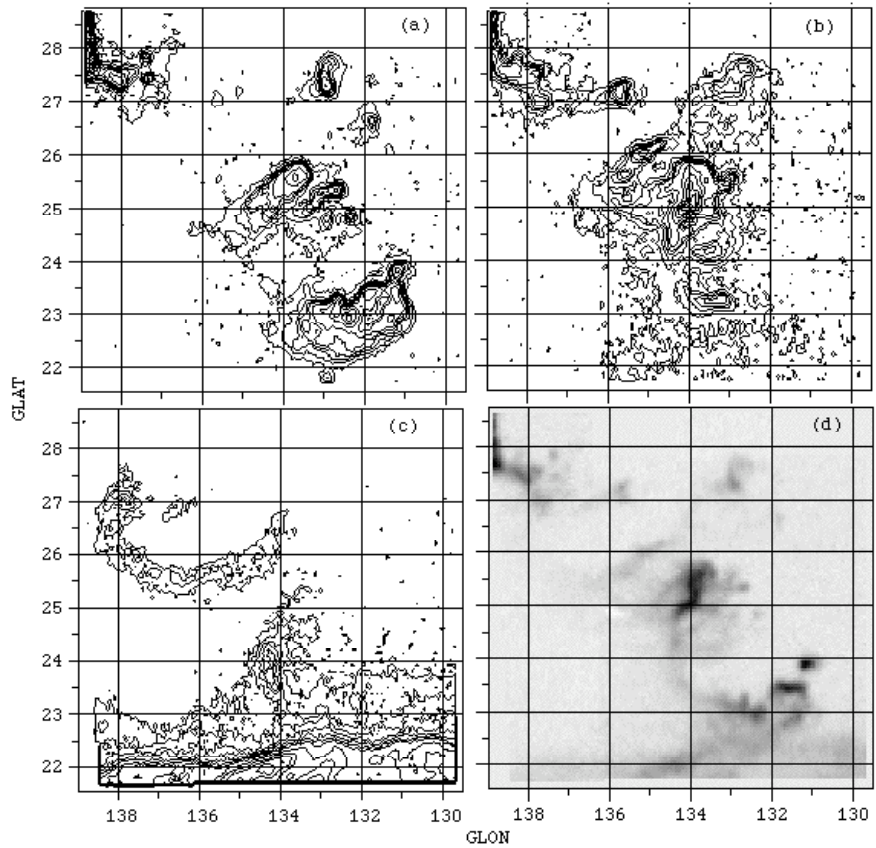}
\caption{The morphology of the anomalous-velocity feature A0 and a small segment of AI (upper left) in galactic coordinates.  (a) From $-$220 to $-$180 km s$^{-1}$. The structure at the bottom of the map, around {\it b}$=$ 23\arcdeg\ is the feature that has been called AO in the literature, see text.  (Contour levels of the integrated antenna temperature are from 1.5 in steps of 3.0 and from 18 at intervals of 10 K.km s$^{-1}$.)  (b) From $-$180 to $-$150 km s$^{-1}$. (Same contour levels.)  (c) From $-$150 to $-$100 km s$^{-1}$ showing a distinct segment of filament, center left, which has been labeled ARC. (Contour levels from 5 in steps of 3 and from 30 in steps of 8 K.km s$^{-1}$. (d) An Inverted Gray-scale representation  of all the emission integrating from $-$220 to $-$120 km s$^{-1}$.  This form of display begins to reveal the twisted filamentary nature of the HI gas. 
}
\end{figure}

\clearpage

\begin{figure}
\figurenum{2}
\epsscale{0.8}
\plotone{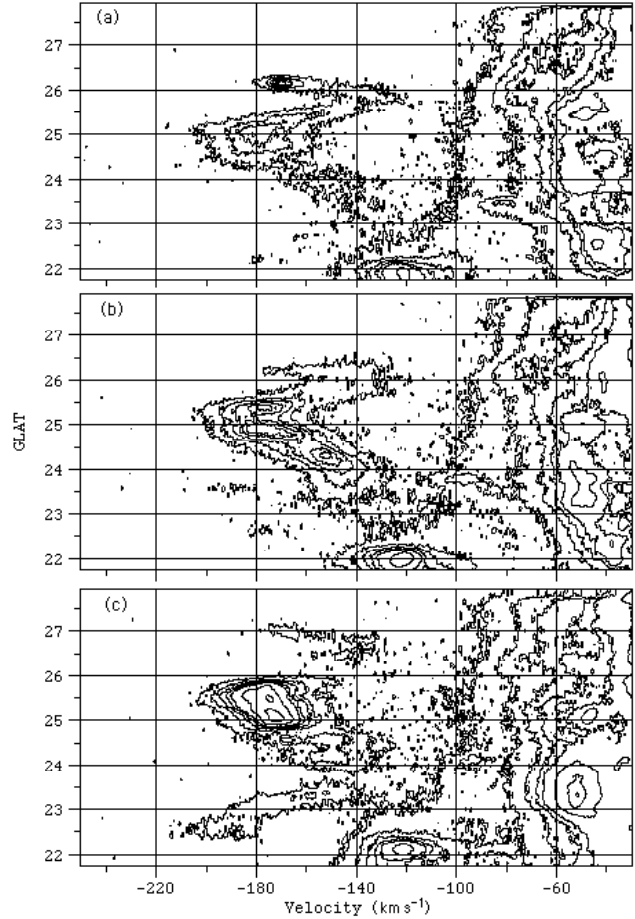}
\caption{The wave-like nature of the HI structure in A0 is evident in these three {\it b,v} diagrams, each of which plots the antenna temperature averaged over three adjacent longitudes.  (a) Centered at {\it l} $=$ 134.\arcdeg79.   (b) Centered at {\it l} $=$ 134.\arcdeg50. (c) Centered at {\it l} $=$ 133.\arcdeg92.  All contour levels are from 0.15 K to 1.15 K in steps of 0.25 K and then from 2 K in steps of 2 K.}
\end{figure}
\clearpage

 \begin{figure}
\figurenum{3}
\epsscale{0.8}
\plotone{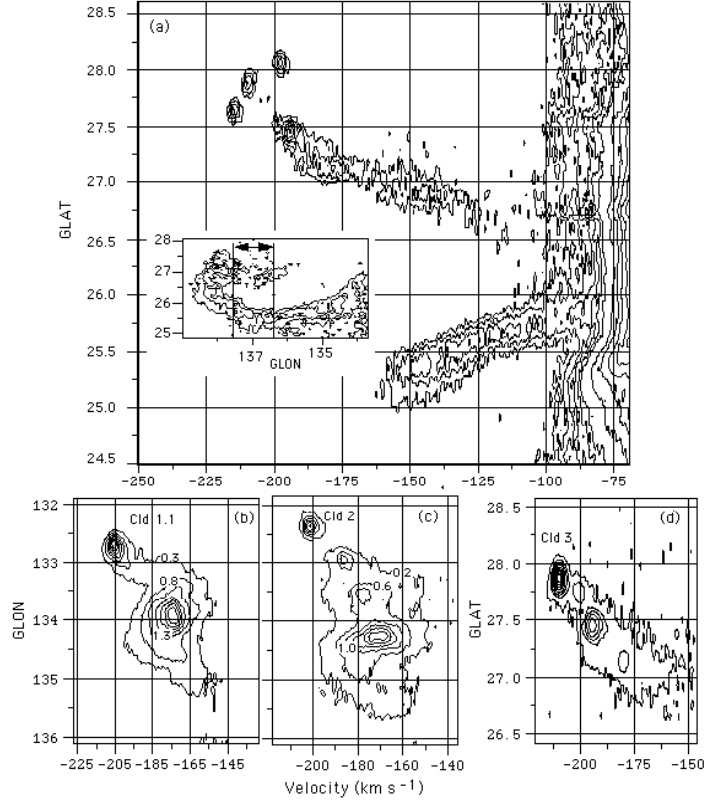}
\caption{(a) The {\it l,v} map averaging the data from {\it l} $=$ 136.\arcdeg3 to 137.\arcdeg3.  Antenna temperature contours are from 0.1 K in steps of 0.05 K and then from 0.5 K in steps of 0.1 K.  The inset shows ARC, taken from Fig. 1c.  The vertical lines marks the longitude range used for the main figure.  The velocity gradient across the ARC seen in the lower part of the main figure suggests rotation about an axis.  At the top of the figure the three sets of closed contours around {\it b} $=$ 28\arcdeg\ include Cloud 3 at the center.  These are located at the edge of AI, see text.  (b) The {\it l,v} map for Cloud 1.1 at {\it b} $=$ 25.\arcdeg33.  Contour levels in K are indicated. (c) The {\it l,v} map for Cloud 2 at {\it b} $=$ 24.\arcdeg85. Contour levels in K are indicated. (d) The {\it b,v} map for Cloud 3.1 at {\it l} $=$ 137.\arcdeg36 $\pm$ 0.\arcdeg0583.  Contour levels are from 0.02 in steps of 0.05 and from 0.19 in steps of 0.1 K. 
}
\end{figure}

\clearpage

 \begin{figure}
\figurenum{4}
\epsscale{0.9}
\plotone{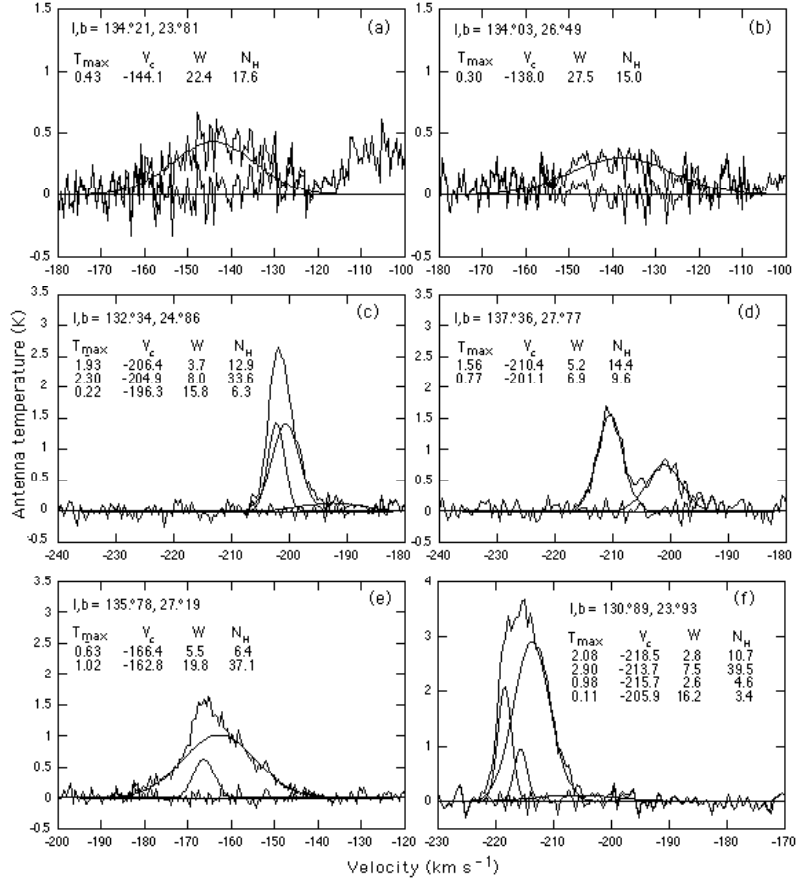}
\caption{Examples of profiles and Gauss fits toward the six directions indicated, together with the parameters that define the profiles.  These are samples toward the features listed in Tables 1 \& 2, as follows: (a) UMB, (b) Cswy B, while (c) $-$ (f) are toward Clouds 2, 3, 5 and PairN respectively.}
\end{figure}

\clearpage

 \begin{figure}
\figurenum{5}
\epsscale{0.7}
\plotone{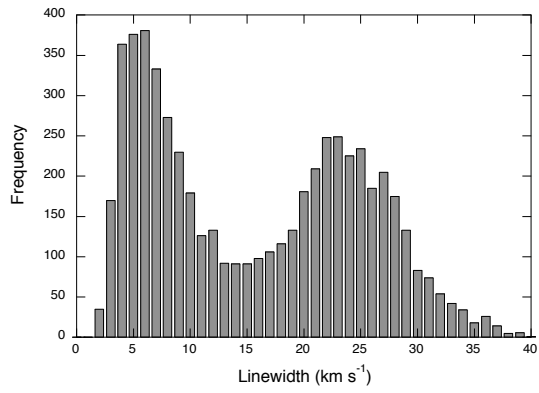}
\caption{Histogram of the nearly 6,000 line widths derived from the Gaussian analysis.  The components with line widths around 24 km s$^{-1}$ are associated with the filament and those around 6 km s$^{-1}$ are associated with the  small diameter clouds.  
}
\end{figure}

\clearpage

 \begin{figure}
\figurenum{6}
\epsscale{1.0}
\plotone{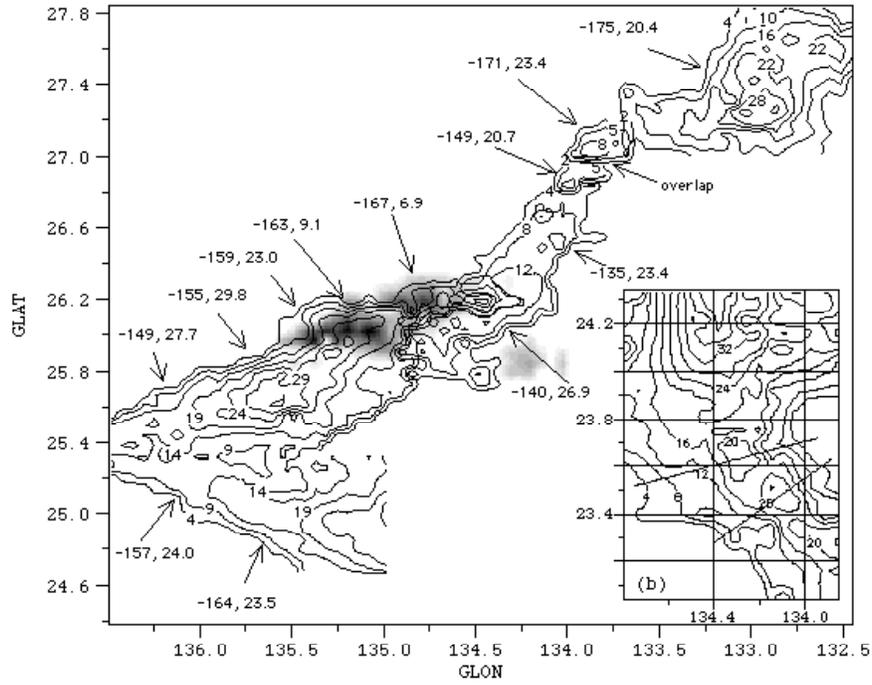}
\caption{A composite of contour maps of the most readily mapped broad line width component, filament features with average center velocities and line widths in km s$^{-1}$ for various segments indicated.  The contour maps are overlain on an inverted gray scale image showing the morphology of a narrow line width cloud component, line width 6.9 km s$^{-1}$, as well as a transition feature of intermediate line width of 9.1 km s$^{-1}$ that together appear to form a bridge between the two broad components.  Labeled contours are column densities in units of 10$^{18}$ cm$^{-2}$.  The inset shows another area of the filament, see text.
}
\end{figure}

\clearpage

 \begin{figure}
\figurenum{7}
\epsscale{0.7}
\plotone{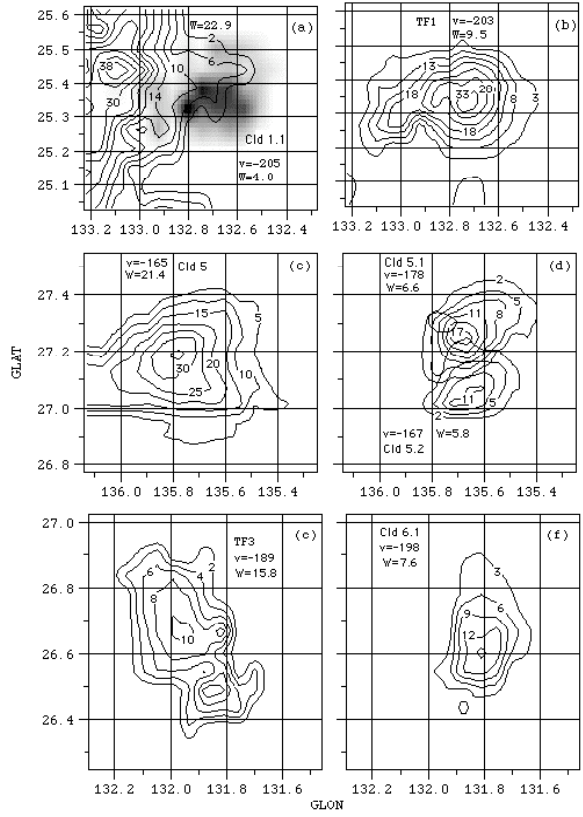}
\caption{Column density maps toward a number of features with contours labeled in units of 10$^{18}$ cm$^{-2}$ and average velocities and line widths indicated. (a) Contours define a broad emission line feature that is part of the central area in Fig. 1d shown together with the inverted gray-scale image of the narrow line width component Cloud 1.1, peak column density 15 x 10$^{18}$ cm$^{-2}$. (b) The column density of transition feature TF1.  (c) The broad line width component that dominates the emission from the Cloud 5 area while (d) shows the morphology of two narrow line width components found for the same area superimposed on one another.  (e) The morphology of another transition line width feature, TF3, exhibiting a large range of line widths. (f) Contours for the associated narrow line width component,  Cloud 6.1.  
}
\end{figure}

\clearpage

\begin{figure}
\figurenum{8}
\epsscale{0.8}
\plotone{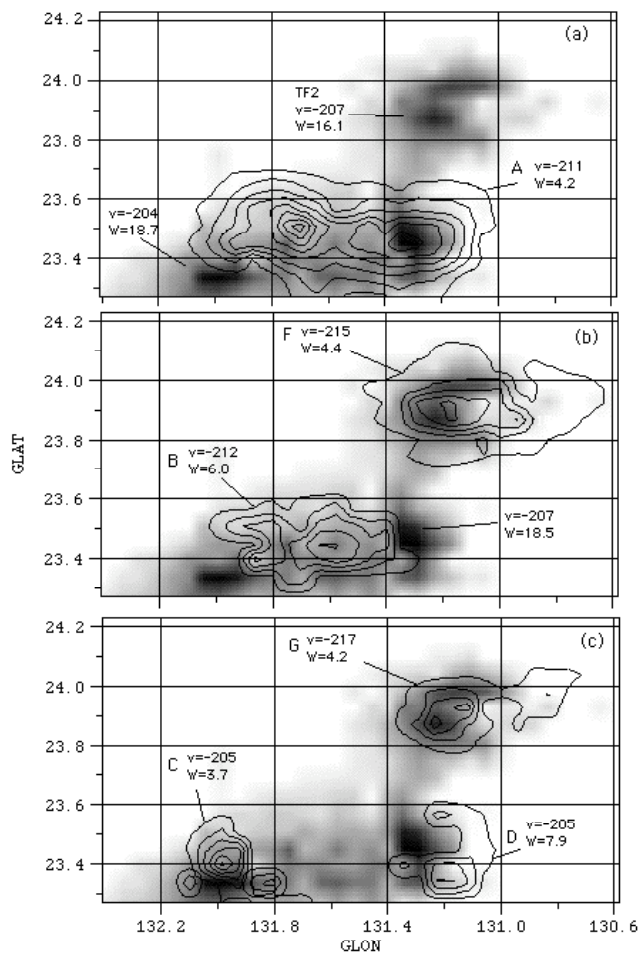}
\caption{(a) Inverted gray scale image of broad line width features with average velocities and line width for small ares indicated. The line widths are characteristically a little narrower than for the majority of the A0 filament segments, see text.  (b) \& (c) Contour maps of the column density of often overlapping narrow line width cloud components, see Table 2, superimposed on the samebroad-line-width, gray-scale image as in (a).  The average velocities and line width for individual features are noted together with the value of the peak HI column densities in units of 10$^{18}$ cm$^{-2}$.
}
\end{figure}

\clearpage

 \begin{figure}
\figurenum{9}
\epsscale{0.8}
\plotone{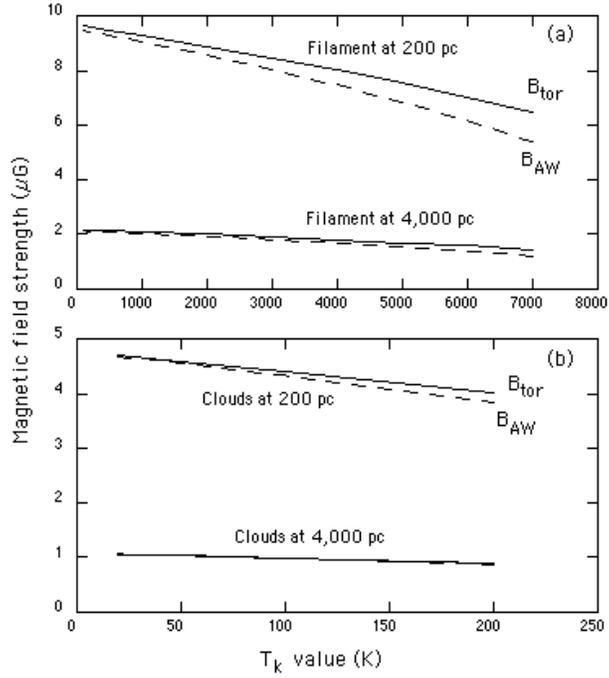}
\caption{ The dependance of the average derived magnetic fields as a function of assumed kinetic temperature of the HI in high-velocty feature A0 for two assumed distances.  (a) For the filament: solid line - toroidal field; dashed line - field derived from the Alfv\`en velocity, see \S6.  (b) The derived fields for the clouds, same notation.
}
\end{figure}

\end{document}